\documentclass[epj,twocolumn]{webofc}
\usepackage[varg]{txfonts}   
\usepackage{graphicx}
\usepackage{color}

\newcommand{\ind}[1]{_{\mathrm{#1}}}
\newcommand{\diff}{\mathrm{d}}
\newcommand\DPi{\Delta\Pi\ind{1}}
\newcommand\Dnu{\Delta\nu}

\newcommand\np{{n\ind{p}}}
\newcommand\ngrav{{n\ind{g}}}
\newcommand\nmax{n\ind{max}}
\newcommand\dl{d_{0\ell}}
\newcommand\epsp{\varepsilon\ind{p}}
\newcommand\epsg{\varepsilon\ind{g}}
\newcommand\dnurot{\delta\nu\ind{rot}}
\newcommand\dnurotcore{\delta\nu\ind{rot, core}}
\newcommand\dnurotenv{\delta\nu\ind{rot, env}}

\newcommand\nup{\nu\ind{p}}
\newcommand\nug{\nu\ind{g}}

\newcommand\Dtaum{\Delta\tau\ind{m}}
\newcommand\xrot{x\ind{rot}}
\newcommand\numax{\nu\ind{max}}

\wocname{epj}
\woctitle{Seismology of the Sun and the Distant Stars 2016}
\begin{document}
\title{Rapidly rotating red giants}
%

\author{\firstname{Charlotte} \lastname{Gehan}\inst{1}\thanks{\email{charlotte.gehan@obspm.fr}} \and
        \firstname{Benoît} \lastname{Mosser}\inst{1} \and
        \firstname{Eric} \lastname{Michel}\inst{1}
}

\institute{Observatoire de Paris, LESIA, Université Pierre et Marie Curie, Université Paris Diderot, PSL}

\abstract{%
Stellar oscillations give seismic information on the internal properties of stars. Red giants are targets of interest since they present mixed modes, wich behave as pressure modes in the convective envelope and as gravity modes in the radiative core. Mixed modes thus directly probe red giant cores, and allow in particular the study of their mean core rotation.\\
The high-quality data obtained by CoRoT and \textit{Kepler} satellites represent an unprecedented perspective to obtain thousands of measurements of red giant core rotation, in order to improve our understanding of stellar physics in deep stellar interiors.\\
We developed an automated method to obtain such core rotation measurements and validated it for stars on the red giant branch. In this work, we particularly focus on the specific application of this method to red giants having a rapid core rotation. They show complex spectra where it is tricky to disentangle rotational splittings from mixed-mode period spacings. We demonstrate that the method based on the identification of mode crossings is precise and efficient. The determination of the mean core rotation directly derives from the precise measurement of the asymptotic period spacing $\DPi$ and of the frequency at which the crossing of the rotational components is observed.
}
\maketitle
%
\section{Introduction}
\label{intro}
Red giant stars are known to exhibit solar-like oscillations, corresponding mostly to pressure modes resulting from acoustic waves stochastically excited by turbulent convection in the external envelope. These low-frequency oscillations are described by the universal red giant oscillation pattern \cite{Mosser_2013}. In combination with effective temperatures, global seismic parameters of pressure modes provide precise estimates of stellar masses and radii which are independent of modelling \cite{Kallinger}.\\
However, the space based CoRoT and \textit{Kepler} missions have revealed that red giants also present mixed modes, which behave as pressure modes in the convective envelope and as gravity modes in the radiative interior \cite{Beck_2011}. Mixed modes thus probe the physical conditions in stellar interiors. In particular, dipole mixed modes allow us to identify the evolutionary stage of evolved stars: it is now possible to distinguish core-helium burning stars from hydrogen-shell burning stars \cite{Mosser_2014}. They moreover provide the measurement of the asymptotic period spacing $\DPi$ \cite{Mosser_2012b} which is linked to the size of the core \cite{Montalban}. Their observation also gives access to the differential-rotation profile in red giants \cite{Beck}.\\
Rotation strongly impacts stellar structure and evolution \cite{Lagarde}. The measurement of the mean core rotation of about 300 red giants analyzed by \cite{Mosser_2012c} has revealed that a very efficient angular momentum transport process from the core to the envelope is at work in red giants: the mean core rotation slows down as stars evolve along the red giant branch (RGB) while their core is contracting.
It is thus crucial to study the evolution of the core rotation for a set of red giants as large as possible in order to reach a deep understanding of the physical mechanisms governing the most inner regions of stars.\\
The era of massive high-precision photometric data from space initiated by CoRoT and \textit{Kepler} satellites opens the way to thousands of measurements of red giant core rotation. We aim at developing a method allowing an efficient, rapid and automated measurement of red giant core rotation, based on the stretching of the spectra in order to disentangle the complex mixed-mode spectra. In this work we focus in particular on red giants with a rapid core rotation, where the rotational splittings are larger than half the mixed-mode spacing and complicate the interpretation of the spectra.


\section{Disentangling red giant spectra in the case of rapidly rotating red giants}
\label{sec-1}
\begin{figure}[t]
\centering
\includegraphics[width=\hsize,clip]{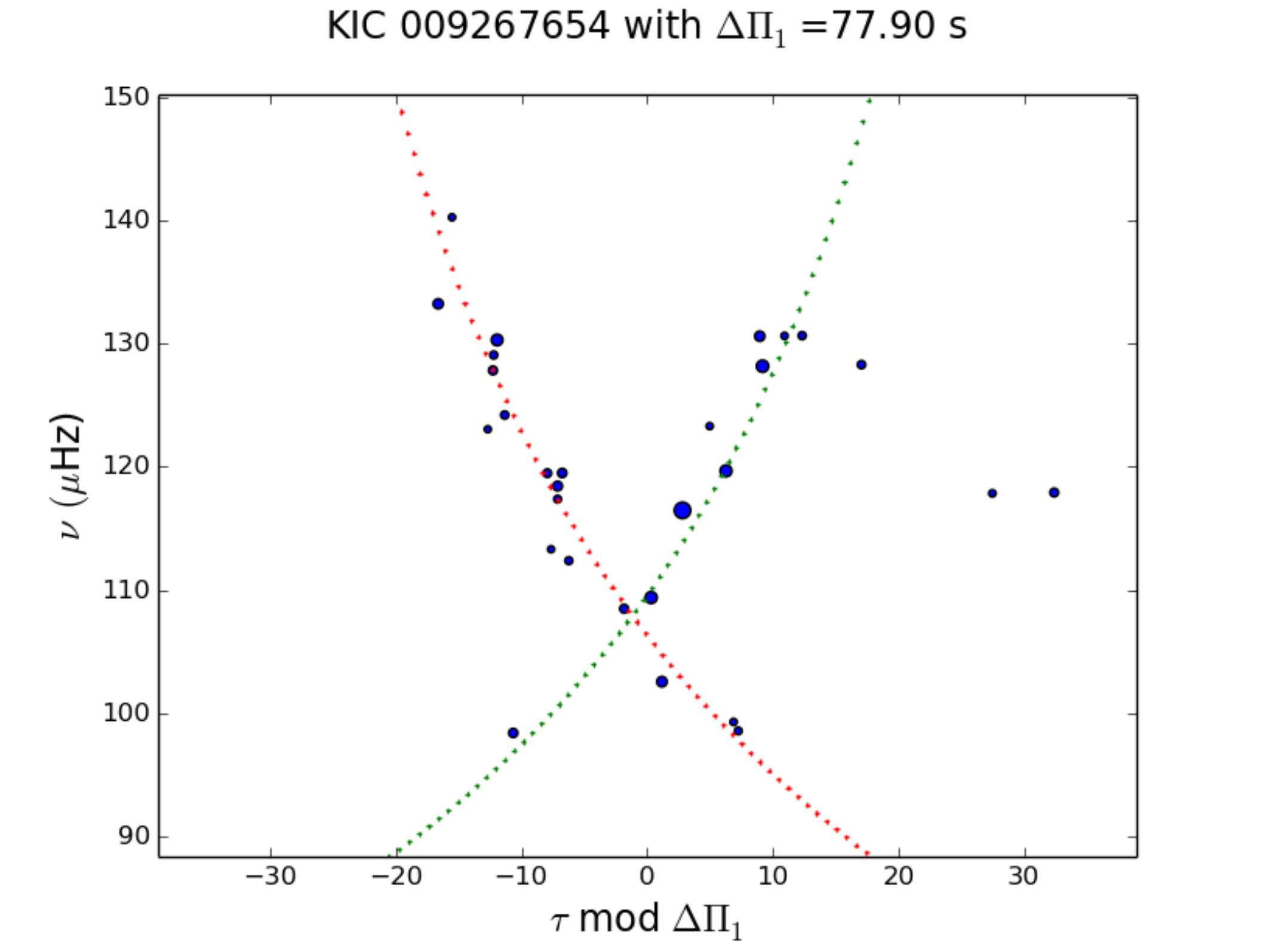}
\caption{Échelle diagram for KIC 9267654 star with $\DPi = 77.9$ s. The rotational multiplet component corresponding to the azimuthal order $m = + \, 1$ is identified in red, and the rotational multiplet component corresponding to the azimuthal order $m = - \, 1$ is identified in green.}
\label{fig-1}       
\end{figure}

When the rotational splitting exceeds half the mixed-mode period spacing, it is very difficult to disentangle the rotational multiplet components from the mixed modes. A part of the difficulty comes from the uneven mixed-mode pattern: pressure dominated mixed-modes are nearly equally spaced in frequency, with a spacing close to the large separation $\Dnu$, whereas gravity dominated mixed-modes are nearly equally spaced in period, with a spacing close to the gravity mode-period spacing $\DPi$ \cite{Mosser_2012b}.\\
In the Astro Fluid 2016 Conference Proceedings edited by EAS Publications Series (in press), we present in detail the method we developed in order to obtain automated measurements of the mean core rotation of red giants, and validate it for stars on the RGB, for which automated measurements are fully consistent with manual measurements \cite{Mosser_2012b}. The principle of the method is based on the stretching of frequency spectra, which ensures that mixed modes are regularly spaced. The frequency is changed into
\begin{equation}
\diff \tau = \frac{\diff \nu}{\zeta \nu^2},
\end{equation}
where $\tau$ is the stretched period \cite{Mosser_2015}, and $\zeta$ is defined by
\begin{equation}
\zeta = \left[1 + \frac{1}{q} \frac{\nu^2 \DPi}{\Dnu} \frac{\cos^2 \left[\pi \frac{1}{\DPi} \left(\frac{1}{\nu} - \frac{1}{\nug} \right)\right]}{\cos^2 \left(\pi \frac{\nu - \nup}{\Dnu}\right)} \right]^{-1},
\end{equation}
where $q$ is the coupling parameter of mixed modes, $\DPi$ is the gravity mode period spacing, $\Dnu$ is the large separation, $\nug$ are the pure dipole gravity mode frequencies, $\nup$ are the pure pressure mode frequencies.\\
For the pure dipole gravity mode frequencies $\nug$ we can use the first-order asymptotic expansion \cite{Tassoul}
\begin{equation}
\frac{1}{\nug} = -(\ngrav + \epsg) \, \DPi,
\end{equation}
where $\ngrav$ is the gravity radial order usually defined as a negative value, and $\epsg$ is a small but complicated function sensitive to the stratification near the boundary between the radiative
core and the convective envelope.\\
For the pure pressure mode frequencies $\nup$ we use the universal red giant oscillation pattern \cite{Mosser_2013}
\begin{equation}
\nup = \left(\np + \frac{1}{2} + \epsp + \dl + \frac{\alpha}{2} [\np - \nmax]^2 \right) \Dnu,
\end{equation}
where $\np$ is the pressure radial order, $\epsp$ is the pressure mode offset, $\dl$ is the small separation, $\alpha$ represents the curvature of the oscillation pattern, and $\nmax = \numax / \Dnu - \epsp$ is the non-integer order at the frequency $\numax$ of maximum oscillation signal.\\
In the stretched spectra, rotation induces a small departure from an evenly spaced pattern. The stretched period spacing between rotational multiplet components with the same azimuthal order $m$ expresses \cite{Mosser_2015}
\begin{equation}
\Dtaum = \DPi (1 + m \xrot)
\label{Dtaum}
\end{equation}
It presents a small departure from $\DPi$, expressed by
\begin{equation}
\xrot = 2 \, \zeta \, \frac{\dnurot}{\nu},
\label{xrot}
\end{equation}
where $\dnurot$ is the rotational splitting and $\numax$ is the frequency of maximum oscillation signal.\\
It is then possible to build échelle diagrams based on this stretched period, where the different components of the rotational multiplet draw ridges and are disentangled. Hence, it is easy to obtain measurements of the mean core rotation.\\
In the present work, we examine the case of stars with a large rotational splitting, which complicates the frequency spectra pattern.
In échelle diagrams representing $\nu$ as a function of $\tau$ modulo $\DPi$, a crossing of the different rotational multiplet components occurs when the rotational splitting is a multiple of half the mixed-mode frequency spacing. Figure~\ref{fig-1} represents the échelle diagram of the rapid rotator KIC 9267654. This star is seen equator-on so only the two rotational multiplet components with the azimuthal orders $m = \pm \, 1$ are visible. We clearly identify a crossing of the components at the frequency $\nu = 108$ $\mu$Hz.

\begin{figure}[t]
\centering
\includegraphics[width=\hsize,clip]{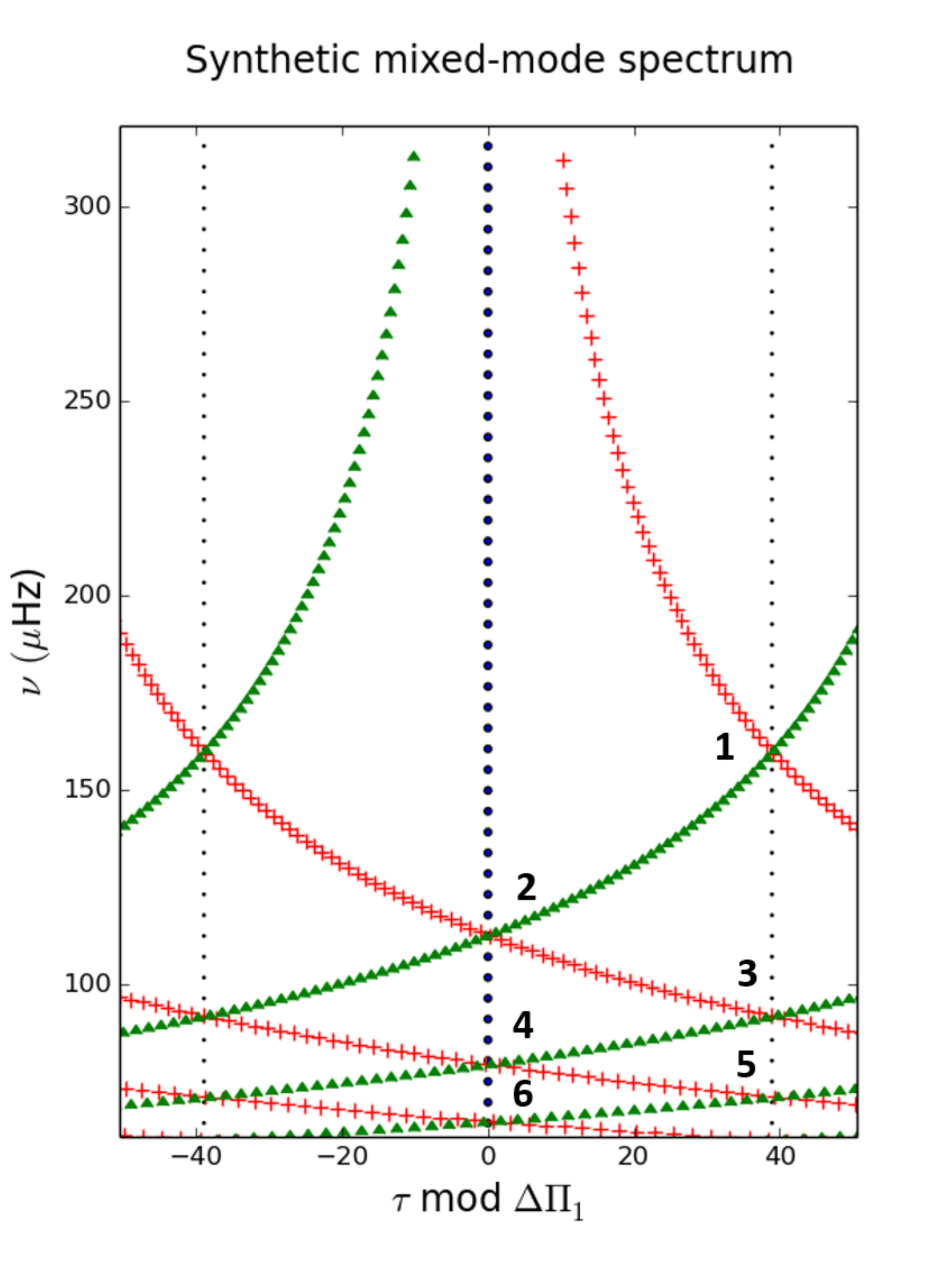}
\caption{Échelle diagram obtained with Equations \eqref{Dtaum} and \eqref{xrot} for a synthetic mixed-mode spectrum with $\DPi = 77.9$ s and $\dnurot = 1$ $\mu$Hz. The rotational multiplet components corresponding to the azimuthal orders $m = \lbrace -1, 0, 1 \rbrace$ are respectively represented by green triangles, large blue dots and red crosses. Vertical black dotted lines represent the $\tau = \left(\pm \, \DPi / 2 \right)$ mod $\DPi$ values. The crossing orders are identified by black numbers.}
\label{fig-2}       
\end{figure}






\section{Crossing of rotational multiplet components}
\label{sec-2}

We aim at deriving a relation between the rotational splitting and the mixed-mode spacing which expresses the conditions leading to a crossing of the different multiplet components.
The condition to have a crossing of the multiplet components is
\begin{equation}
\dnurot = k \, \frac{\delta \nu}{2},
\label{crossing}
\end{equation}
where $\delta \nu$ is the mixed-mode frequency spacing, and $k$ is a positive integer.\\
The rotational splitting expresses \cite{Goupil}
\begin{equation}
\dnurot = \zeta \, \dnurotcore + (1-\zeta) \, \dnurotenv,
\label{dnurot_ini}
\end{equation}
where $\dnurotcore$ and $\dnurotenv$ are the rotational splittings that represent respectively the mean rotation of the radiative core and the mean rotation of the convective envelope.\\
Red giants have a radiative core rotating much faster than the envelope. When the surface rotation can be neglected, we can rewrite Equation \eqref{dnurot_ini} as
\begin{equation}
\dnurot = \zeta \, \dnurotcore,
\label{dnurot}
\end{equation}
which means that the rotational splittings associated to the core rotation are modulated along the oscillation spectra by the $\zeta$ function.\\
The period spacing between consecutive mixed-modes expresses \cite{Mosser_2015}
\begin{equation}
\Delta P = \zeta \, \DPi.
\end{equation}
The relation between $P$ and $\nu$ expresses
\begin{equation}
\frac{\diff P}{P} = - \frac{\diff \nu}{\nu}.
\end{equation}
Hence the frequency spacing $\delta \nu$ between consecutive mixed-modes expresses
\begin{equation}
\delta \nu = \nu^2 \, \Delta P.
\label{dnu}
\end{equation}
Thus we can rewrite Equation \eqref{crossing} with Equation \eqref{dnurot} and Equation \eqref{dnu} to obtain the condition to have a crossing of the rotational multiplet components. This crossing at frequency $\nu \ind{k}$ is linked to the global seismic properties $\dnurotcore$ and $\DPi$ by
\begin{equation}
\zeta \dnurotcore = k \, \frac{\nu^2 \ind{k} \, \zeta \, \DPi}{2},
\end{equation}
where the $\nu \ind{k}$ are the frequencies where a crossing of the rotational multiplet components occurs at the order $k$. This relation simplifies into
\begin{equation}
\dnurotcore = k \, \frac{\nu^2 \ind{k} \, \DPi}{2}.
\label{dnurotcore}
\end{equation}
We note that the $\zeta$ function is not involved in the measurement of the rotational splitting. The $\DPi$ parameter has been measured for about 5000 stars \cite{Vrard}. We just need to identify in échelle diagrams one of the frequencies $\nu \ind{k}$ corresponding to a crossing of the multiplet components and the order of the crossing $k$ in order to obtain a precise estimate of the core rotation of these rapidly rotating red giants. 

\section{Measurement of the rotational splitting}
\label{sec-3}

The next step consists in identifying the order $k$ of the crossing of the rotational multiplet components.

\subsection{Crossings of the rotational multiplet components in a synthetic spectrum}
\label{sec-3-bis}

Figure~\ref{fig-2} represents an échelle diagram for a synthetic mixed-mode spectrum based on Equations \eqref{Dtaum} and \eqref{xrot}. We note that several crossings of the multiplet components occur at different frequencies. The order of the crossing decreases with frequency. At high frequency, the multiplet components are almost parallel and no crossing is possible. We note that the frequency spacing of the crossings increases with frequency. The parity of the order $k$ modifies the pattern drawn by the multiplet components in the échelle diagram. When $k$ is an even number the three components corresponding to the azimuthal order values $m = \lbrace -1, 0, 1 \rbrace$ overlap at $\tau$ mod $\DPi = 0$, while when $k$ is an odd number only the $m = \pm \, 1$ components overlap at $\tau = \left(\pm \, \DPi / 2 \right)$ mod $\DPi$.\\
Usually, the number of crossings in the range where mixed modes are excited is low and only one crossing is visible, as it is the case for KIC 9267654 star. The existence or the absence of possible crossings at $\tau = \left(\pm \, \DPi / 2 \right)$ mod $\DPi$  allows the identification of the crossing order $k$.

\subsection{Identification of the order of the observed crossing}
\label{sec-3-ter}

The identification of the order $k$ of the crossing of the rotational multiplet components is achieved by computing the frequencies where crossings occur. These frequencies are given by
\begin{equation}
\nu \ind{k} = \sqrt{\frac{2 \, \dnurotcore}{k \, \DPi}}.
\end{equation}
In order to correctly span the possibilites, the different possible $k$ values are explored for a fixed $\dnurotcore$ value obtained via Equation \eqref{dnurotcore}.
Then the frequencies $\nu \ind{k}$ which match the observations provide the value of the order $k$.
An example is given in Table~\ref{tab-1} for KIC 9267654 star. For this star, the order of the crossing is most likely $k = 1$ and the corresponding rotational slitting value $\dnurotcore  = 454$ nHz, since no crossing occurs for frequencies greater than 108 $\mu$Hz.\\
Table~\ref{tab-2} gives the measured rotational splitings for other stars presenting overlaping rotational multiplet components. We note that the identication of the frequency where multiplet components overlap leads to the measurement of high core rotation rates as large as 5 $\mu$Hz, as it is the case for KIC 9227589.

\begin{table}
\centering
\caption{Determination of the order $k$ of the crossing}
\label{tab-1}       
\begin{tabular}{ccccc}
\hline
$k$ & $\nu \ind{k-1}$ & $\nu \ind{k}$ & $\nu \ind{k+1}$ & $\dnurotcore$\\
 & \small{($\mu$Hz)} & \small{($\mu$Hz)} & \small{($\mu$Hz)} & \small{(nHz)}\\\hline
1 & - & 108.0 & 76.4 & 454 \\
2 & 152.7 & 108.0 & 88.2  & 909 \\
3 & 132.3 & 108.0 & 93.5  & 1363 \\
4 & 124.7 & 108.0 & 96.6 & 1817 \\\hline
\end{tabular}
\begin{center}
\small{Order $k$ of the observed crossing, frequencies $\lbrace \nu \ind{k-1}, \nu \ind{k}, \nu \ind{k+1} \rbrace$ where a crossing occurs and associated $\dnurotcore$ values for KIC 9267654 star having $\DPi = 77.9$ s.}
\end{center}
\end{table}

\begin{table}
\centering
\caption{Measurement of rotational splittings}
\label{tab-2}       
\begin{tabular}{ccccccc}
\hline
KIC & $k$ & $\DPi$ & $\nu \ind{k-1}$ & $\nu \ind{k}$ & $\nu \ind{k+1}$ & $\dnurotcore$\\
 &  & \small{(s)} & \small{($\mu$Hz)} & \small{($\mu$Hz)} & \small{($\mu$Hz)} & \small{(nHz)}\\\hline
5380775 & 3 & 78.20 & 91.9 & 75.0 & 65.0 & 660 \\
6949009 & 2 & 72.79 & 113.1 & 80.0 & 65.3 & 466 \\
7257175 & 2 & 72.60 & 134.4 & 95.0 &  77.6 & 655 \\
7267119 & 3 & 73.90 & 98.0 & 80.0 & 69.3 & 709 \\
9227589 & 4 & 76.00 & 207.8 & 180.0 & 161.0 & 4925 \\
10387370 & 3 & 70.80 & 89.4 & 73.0 & 63.2 & 566 \\\hline
\end{tabular}
\end{table}

\section{Uncertainties}
\label{sec-4}

The uncertainty on $\dnurot$ expresses
\begin{equation}
\frac{\delta (\dnurotcore)}{\dnurotcore} = \sqrt{\left[\frac{\delta (\DPi)}{\DPi}\right]^2 + 4 \, \left[\frac{\delta \nu \ind{k}}{\nu \ind{k}}\right]^2}
\end{equation}
where $\delta (\DPi)$ and $\delta \nu \ind{k}$ are respectively the uncertainties on $\DPi$ and on $\nu \ind{k}$.\\
Typical values for the relative uncertainties are $\delta (\DPi) / \DPi \simeq 1\%$ and $\delta \nu \ind{k} / \nu \ind{k} \simeq 1\%$. Thus the relative uncertainty on $\dnurotcore$ is $\delta (\dnurotcore) / \dnurotcore \simeq 3 \%$.

\section{Conclusions}
\label{sec-con}
By correcting frequency spectra into stretched period spectra where mixed modes are regularly spaced, we were able to disentangle intricated red giant spectra where the rotational splitting exceeds half the mixed-mode period spacing. These large rotational splittings are the signature of large core rotation values, which induce crossings of the rotational multiplet components in stretched period échelle diagrams. This method succeeds in measuring precise high core rotation rates with the knowledge of the gravity mode period spacing $\DPi$, by identifying the frequency corresponding to a crossing of the multiplet components. The relative uncertainty obtained on the rotational splitting measurement is around 3\%.

 \bibliography{GEHANCharlotte.bbl}

\end{document}